\documentclass[10pt,conference,letterpaper]{IEEEtran}
\IEEEoverridecommandlockouts

\usepackage{amsmath,amssymb,amsthm}
\usepackage{siunitx}
\usepackage{mathtools}
\usepackage{graphicx}
\usepackage{caption}
\usepackage{subcaption}
\usepackage{float}
\usepackage{booktabs}
\usepackage{multirow}
\usepackage{tabularx}
\usepackage{colortbl}
\usepackage{xcolor}
\definecolor{codeblue}{HTML}{355C7D}
\definecolor{codepink}{HTML}{F67280}
\definecolor{codegray}{HTML}{4C4C4C}
\definecolor{codelightblue}{HTML}{6C8EAD}
\definecolor{codelightpink}{HTML}{F8A5AE}
\definecolor{codeoffwhite}{HTML}{F7F7F7}
\definecolor{codedarkpink}{HTML}{E85D75}

\colorlet{steponebg}{codelightblue!20}
\colorlet{steptwobg}{codeoffwhite}
\colorlet{stepthreebg}{codelightpink!28}
\colorlet{boxborder}{codegray!65}
\colorlet{arrowcol}{codeblue!85!codegray}
\colorlet{titlecol}{codegray}

\usepackage{textcomp}
\usepackage{afterpage}
\usepackage{gensymb}
\usepackage[english]{babel}
\usepackage{adjustbox}

\usepackage{enumitem}
\usepackage{array}
\newcolumntype{L}[1]{>{\raggedright\arraybackslash}m{#1}}
\newcolumntype{C}[1]{>{\centering\arraybackslash}m{#1}}
\newcolumntype{R}[1]{>{\raggedleft\arraybackslash}m{#1}}
\newsavebox{\myboxtable}
\usepackage{tikz}
\usetikzlibrary{positioning, calc, fit, backgrounds}

\usepackage{algorithm}
\usepackage[noend]{algpseudocode}
\algrenewcommand\algorithmicrequire{\textbf{Require:}}
\algrenewcommand\algorithmicensure{\textbf{Ensure:}}

\usepackage{cite}

\newcommand{\boldtheorem}[1]{\textbf{#1}}
\newtheorem{theorem}{\boldtheorem{Theorem}}
\newtheorem{lemma}{\boldtheorem{Lemma}}

\usepackage[hidelinks]{hyperref}

\newtheorem{corollary}{\boldtheorem{Corollary}}

\newsavebox{\leftgridbox}

\hypersetup{
    colorlinks=true,
    linkcolor=black,
    citecolor=blue,
    urlcolor=blue,
    pdfborder={0 0 0}
}

\def\BibTeX{{\rm B\kern-.05em{\sc i\kern-.025em b}\kern-.08em
  T\kern-.1667em\lower.7ex\hbox{E}\kern-.125emX}}

\title{
Cascading Failure Dynamics in Integrated Multimodal Urban Transport Networks
}

\author{%
Jinghua Song\textsuperscript{\rm 1,2}, Yuan Wang\textsuperscript{\rm 1,2}, Zimo Yan\textsuperscript{\rm 3,*}\thanks{*Corresponding author: Zimo Yan (yanzimo20@nudt.edu.cn).}%
\vspace{1.6mm}\\
\fontsize{10}{10}\selectfont\itshape
\textsuperscript{\rm 1}School of Urban Design, Wuhan University, Wuhan, China.\\
\textsuperscript{\rm 2}Hubei Habitat Environment Research Centre of Engineering and Technology, Wuhan, China.\\
\textsuperscript{\rm 3}National University of Defense Technology, Changsha, China.\\
\vspace{1mm}
\fontsize{9}{9}\selectfont\ttfamily
\{sjinghua, 2024282090042\}@whu.edu.cn, yanzimo20@nudt.edu.cn
}

\begin{document}
\maketitle

\begin{abstract}
Modern urban resilience is increasingly challenged by systemic risk in multimodal transport networks, where localized disruptions can spread and impair functionality. Existing studies have advanced pairwise topology, static resilience assessment and higher order modelling, but it remains unclear whether static structure can predict overload driven cascade outcomes. To address this gap, we construct a multilayer model of the Wuhan multimodal public transport network and compare resilience indicators, centrality measures, motif participation scores and dynamic load redistribution simulations. Results show that intermodal integration improves indicators, including connectivity, robustness and relocation opportunities, but does not guarantee lower cascade losses. Degree, betweenness and feed forward loop motif scores are weakly associated with functional recoverability across tolerances. These findings suggest that cascade vulnerability is governed more by dynamic redistribution than by static structural prominence, and that limited motif performance should be treated as model specific evidence rather than a rejection of higher order approaches.
\end{abstract}

\section*{Introduction}

Modern cities depend on tightly coupled infrastructure systems whose interdependence supports daily mobility while also increasing vulnerability to disruption \cite{wei2020study,shi2023review,huang2022analysis}. In such settings, localized shocks may spread through transport services and develop into cascading failures with citywide consequences \cite{jiang2020overview,gong2023empirical,daqing2014spatial,zhao2024cascading}. Despite sustained investment in infrastructure renewal, risk management, and service coordination, improving urban resilience remains a persistent challenge \cite{goldbeck2019resilience}.

A multimodal public transport system is naturally represented as a multilayer or multiplex network, in which each transport mode forms a layer and interlayer edges encode transfer opportunities. Foundational studies show that coupling across layers shapes robustness, diffusion, and dependence, providing the formal basis for treating intermodal integration as an analytically meaningful network property rather than a descriptive feature \cite{kivela2014multilayer,dedomenico2013mathematical}. Empirical studies of urban mobility further show that spatial constraints, transfer behaviour, and modal coupling jointly shape accessibility, congestion, service redundancy, and disruption impacts \cite{barthelemy2011spatial,gallotti2014anatomy,strano2015multiplex,chodrow2016demand}. Within this line of work, the Safe to Fail perspective has encouraged resilience assessment beyond failure prevention alone \cite{ahern2011fail}. Representative multilayer frameworks, such as those developed by \textit{Xu and Chopra}, evaluate preparedness, robustness, and interoperability in integrated transport systems \cite{chopra2016network,aleta2017multilayer,aparicio2022assessing,liang2023research}. However, these approaches mainly capture the structural benefits of integration and do not fully explain whether additional transfer links also create pathways through which disruptions can be redistributed and amplified.

A second line of research examines failure propagation as a dynamic process. Percolation-based studies analyze fragmentation under node or edge removal and show that interdependence can substantially increase fragility \cite{havlin2012catastrophic,buldyrev2010catastrophic,parshani2010interdependent,liu2019asymmetry,schneider2011mitigation}. Load and capacity models instead focus on overload redistribution after disruption \cite{jo2021cascading,sun2022vulnerability,guo2020simulation,luo2019passenger,zhang2019dynamic}, while dynamic simulations represent the temporal evolution of failure through behavioural or state transition processes \cite{zhang2023quantifying,goldbeck2019resilience}. In urban transport, such redistribution is shaped by passenger behaviour: rerouting tends to follow least cost choices consistent with User Equilibrium \cite{sheffi1985urban}, and recent studies link this mechanism directly to cascading failure in public transport \cite{zhang2018cascading}. At the aggregate scale, functional collapse is also related to hysteresis, capacity drop, and the percolation of congested clusters \cite{geroliminis2008existence,ambuhl2021understanding,hamedmoghadam2021percolation}. Together, these studies make clear that transport resilience cannot be inferred from topology alone.

A third line of research asks whether pairwise topology is sufficient to characterize vulnerability in multimodal transport systems. Existing multilayer resilience frameworks, including the work of \textit{Xu and Chopra}, are effective for evaluating interconnectedness and safe to fail performance \cite{xu2022network,xu2021resilient}, but they do not fully explain whether intermodal integration reshapes cascade vulnerability in ways that static indicators cannot capture. One possible reason is that fragile dependencies may reside in motifs and other higher order patterns rather than in nodes and edges alone \cite{he2023link,xu2024higher,xia2022extreme}. More broadly, work on hypergraphs and simplicial complexes shows that group interactions can qualitatively alter spreading thresholds and propagation behaviour \cite{battiston2020networks,iacopini2019simplicial,benson2016higher}. At the same time, the gap between topology and dynamics is not a new concern: prior studies have long emphasized that structural importance alone is insufficient for understanding robustness and cascade behaviour \cite{braha2007statistical,braha2016complexity,braha2020patterns}. Our aim is therefore not to restate that general point, but to provide an empirical case from a real multimodal urban transport network in which both conventional centrality and static motif scores show limited ability to predict cascade outcomes under a load and capacity mechanism. This is practically important because historical disruption records often reflect local exposure, reporting bias, or event-specific operating conditions rather than the structural dependencies that govern system-wide failure.

These gaps motivate two central questions. \textit{Can static structural descriptors, including conventional centrality and motif participation, predict functional cascade outcomes in a real multimodal transport network?} \textit{If their predictive power is limited, what dynamic mechanisms determine when intermodal integration mitigates or intensifies cascading failure?} To address these questions, we construct a multilayer representation of the Wuhan multimodal public transport network and evaluate resilience from three complementary perspectives adapted from the resilience cycle framework \cite{xu2023interconnectedness}. Preparedness is measured by the Gini coefficient of node importance, robustness by the indicator $r_b$ under different attack strategies, and interoperability by the relocation rate $R_\ell$. On this basis, we compare traditional structural metrics, motif participation scores, and dynamic cascade outcomes within a unified empirical framework.

\begin{figure*}[htbp]
    \centering
    \includegraphics[width=\textwidth]{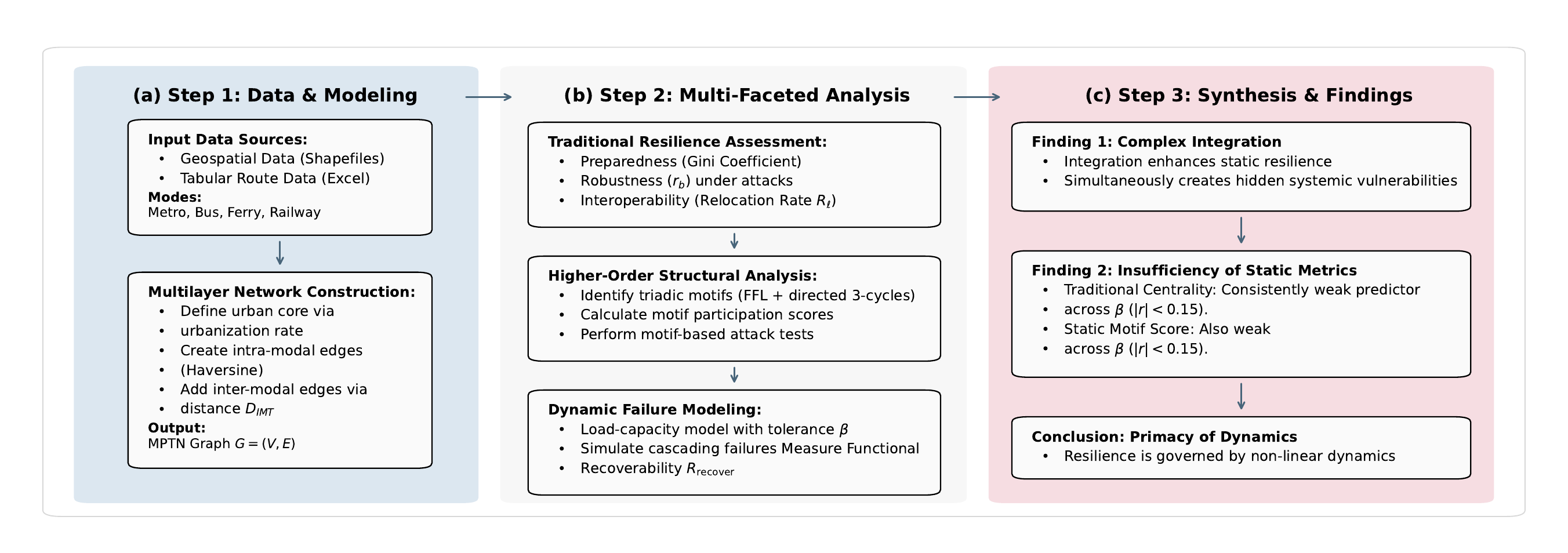}
    \caption{Analytical framework for the Wuhan MPTN. (a) Multilayer network construction. (b) Traditional structural analysis, higher order motif analysis, and dynamic cascade simulation. (c) Synthesis of the main findings: integration improves static resilience indicators, while functional cascade outcomes depend on dynamic redistribution and are weakly predicted by static node descriptors.}
    \label{fig:framework}
\end{figure*}

The main contributions of this paper are threefold: (1) we develop an empirical framework for assessing multimodal transport resilience that combines safe to fail indicators, higher order structural analysis, and load and capacity cascade simulation; (2) we show that intermodal integration improves conventional static resilience indicators, but these gains do not necessarily translate into uniformly lower functional cascade losses; and (3) we provide evidence that static descriptors, including degree, betweenness, and motif participation, have limited predictive value for node-level functional recoverability, which points to the importance of modelling dynamic load redistribution when assessing cascade vulnerability. Figure~\ref{fig:framework} summarizes the overall analytical workflow of this study.

\section*{Results}

This section first establishes subsystem-level baseline context and then reports four empirical findings: (i) static gains from intermodal integration, (ii) dynamic cascade consequences under step-wise integration, (iii) the limited predictive power of static structural descriptors, and (iv) system-level nonlinearity in cascade propagation.

\subsection*{Static effects of integration}

Before evaluating step-wise integration effects, we summarize subsystem-level baselines in the study area. The bus layer is the largest subsystem and provides the strongest relocation capacity, whereas the metro layer is more compact but less spatially substitutable. Ferry and railway remain very small subsystems and should therefore be interpreted primarily as local auxiliary layers rather than structurally dominant components (Table~\ref{tab:network_comparison}).

\begin{table}[t]
\centering
\captionsetup{
    labelsep=colon,
    labelfont=bf,
    textfont=bf,
    justification=raggedright,
    singlelinecheck=false
}
\caption{Subsystem baseline metrics in the study area.}
\label{tab:network_comparison}
\footnotesize
\renewcommand{\arraystretch}{1.08}
\definecolor{tablegray}{HTML}{EAEBEB}

\resizebox{\columnwidth}{!}{%
\begin{tabular}{lrrrr}
\toprule
\rowcolor{tablegray}
Metric & Metro & Bus & Ferry & Railway \\
\midrule
$|V|$ & 179 & 2304 & 3 & 3 \\
$|E|$ & 388 & 5838 & 4 & 4 \\
$S_0$ & 0.905 & 0.991 & 1.000 & 1.000 \\
$\langle l \rangle$ & 10.89 & 21.06 & 1.33 & 1.33 \\
$E$ & 0.1313 & 0.0678 & 0.8333 & 0.8333 \\
$E_{\mathrm{geospatial}}$ & 0.7748 & 0.7819 & 0.8116 & 0.8270 \\
Gini (BC) & 0.483 & 0.728 & 0.667 & 0.667 \\
$r_b$ (Random) & 0.2091 & 0.2614 & 0.4622 & 0.4600 \\
$r_b$ (ND-targeted) & 0.0768 & 0.1205 & 0.3889 & 0.3889 \\
$r_b$ (BC-targeted) & 0.1239 & 0.1991 & 0.3889 & 0.3889 \\
$R_\ell$ ($d_{\max}=750$m) & 0.0122 & 0.4881 & 0.0000 & 0.0000 \\
$R_\ell$ ($d_{\max}=1600$m) & 0.3008 & 0.7483 & 0.0000 & 0.0000 \\
\bottomrule
\end{tabular}%
}

\vspace{2pt}
\begin{minipage}{\columnwidth}
\footnotesize
Note: Metric definitions are given in the main manuscript.
\end{minipage}
\end{table}

\begin{table}[t]
\centering
\captionsetup{labelsep=colon, labelfont=bf, textfont=bf}
\caption{Key static metrics across integration steps under isolated and interconnected configurations.}
\label{tab:mptn_property}
\footnotesize
\renewcommand{\arraystretch}{1.08}
\definecolor{tablegray}{HTML}{EAEBEB}

\begin{tabular*}{\columnwidth}{@{\extracolsep{\fill}}lcccc@{}}
\toprule
Metric & S1 & S2 & S3 & S4 \\
\midrule
\rowcolor{tablegray}
\multicolumn{5}{@{}l}{\textbf{Iso. ($D_{\mathrm{IMT}}=0$ m)}} \\
IMT edges & 0 & 0 & 0 & 0 \\
$S_0$ & 0.905 & 0.920 & 0.919 & 0.918 \\
$\langle l \rangle$ & 10.89 & 21.06 & 21.06 & 21.06 \\
$E$ & 0.1313 & 0.0678 & 0.0678 & 0.0678 \\
$E_{\mathrm{geo}}$ & 0.7748 & 0.7819 & 0.7819 & 0.7819 \\
Gini(BC) & 0.483 & 0.728 & 0.728 & 0.728 \\
$R_\ell(750)$ & 0.0122 & 0.4538 & 0.4532 & 0.4527 \\
$R_\ell(1600)$ & 0.3008 & 0.7160 & 0.7151 & 0.7143 \\
\midrule
\rowcolor{tablegray}
\multicolumn{5}{@{}l}{\textbf{Conn. ($D_{\mathrm{IMT}}=100$ m)}} \\
IMT edges & 0 & 166 & 166 & 168 \\
$S_0$ & 0.905 & 0.992 & 0.991 & 0.991 \\
$\langle l \rangle$ & 10.89 & 19.64 & 19.64 & 19.64 \\
$E$ & 0.1313 & 0.0707 & 0.0707 & 0.0707 \\
$E_{\mathrm{geo}}$ & 0.7748 & 0.7897 & 0.7897 & 0.7889 \\
Gini(BC) & 0.483 & 0.722 & 0.722 & 0.722 \\
$R_\ell(750)$ & 0.0122 & 0.4973 & 0.4967 & 0.4965 \\
$R_\ell(1600)$ & 0.3008 & 0.7454 & 0.7445 & 0.7440 \\
\bottomrule
\end{tabular*}

\vspace{2pt}
\begin{minipage}{\columnwidth}
\footnotesize
Note: S1--S4 denote Step 1 (+Metro), Step 2 (+Bus), Step 3 (+Ferry), and Step 4 (+Railway). Full metric definitions are given in Methods.

\end{minipage}
\end{table}

\begin{figure*}[htbp]
    \centering
    \includegraphics[width=\textwidth]{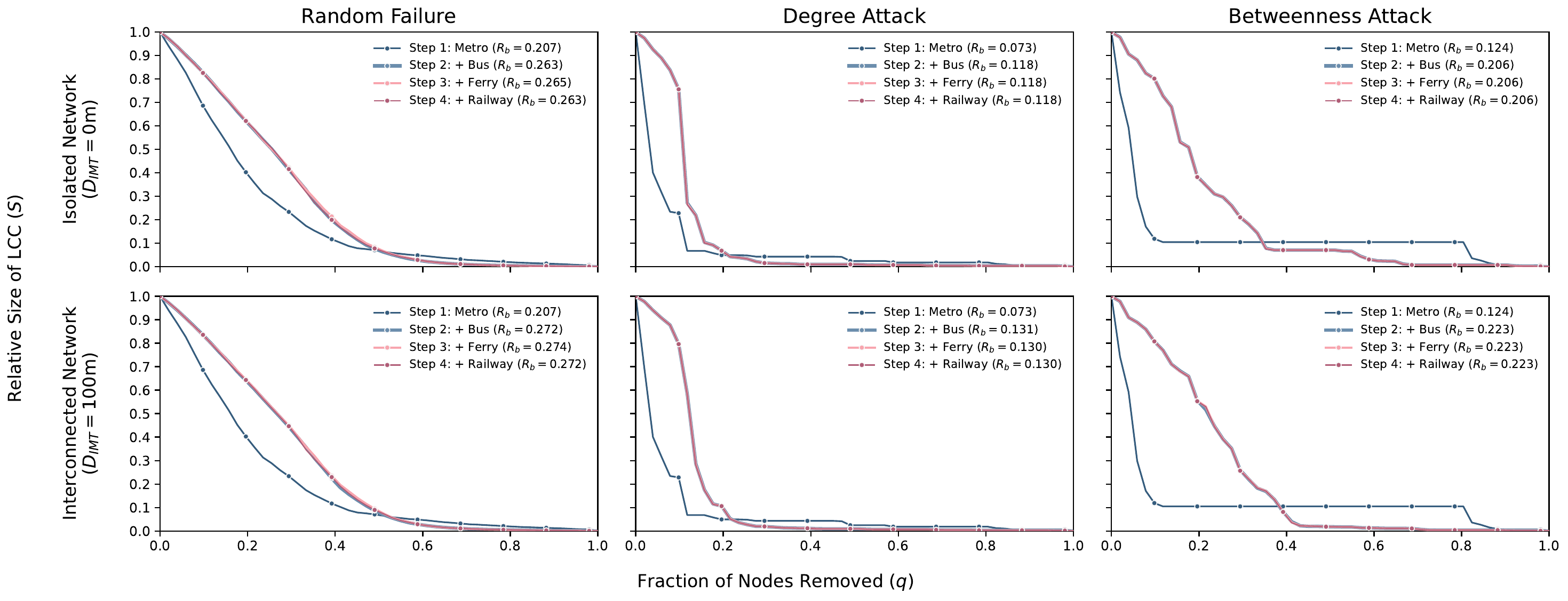}
    \caption{Robustness of the MPTN across integration steps under isolated ($D_{\mathrm{IMT}}=0$ m) and interconnected ($D_{\mathrm{IMT}}=100$ m) configurations.}
    \label{fig:network_robustness_steps}
\end{figure*}

Against this subsystem baseline, intermodal integration improves standard static resilience indicators, with the main change occurring when the bus layer is added. At Step~2, the interconnected network has a larger weakly connected component, a shorter mean path length, slightly higher global efficiency, and higher relocation rates than the isolated network (Table~\ref{tab:mptn_property}). The robustness curves under random, degree-targeted, and betweenness-targeted attacks show the same pattern (Fig.~\ref{fig:network_robustness_steps}).

Changes after Step~2 are small, indicating that metro--bus coupling accounts for most of the structural gain, whereas adding ferry and railway contributes little further improvement. Under static criteria, integration thus improves connectivity, efficiency, interoperability, and robustness.

Mode-level relocation diagnostics show that these gains are concentrated in the metro layer, while the bus layer changes only marginally after interconnection (Table~\ref{tab:consolidated_rl}). Static benefits are therefore unevenly distributed across modes.

\begin{figure*}[htbp]
    \centering
    \includegraphics[width=\textwidth]{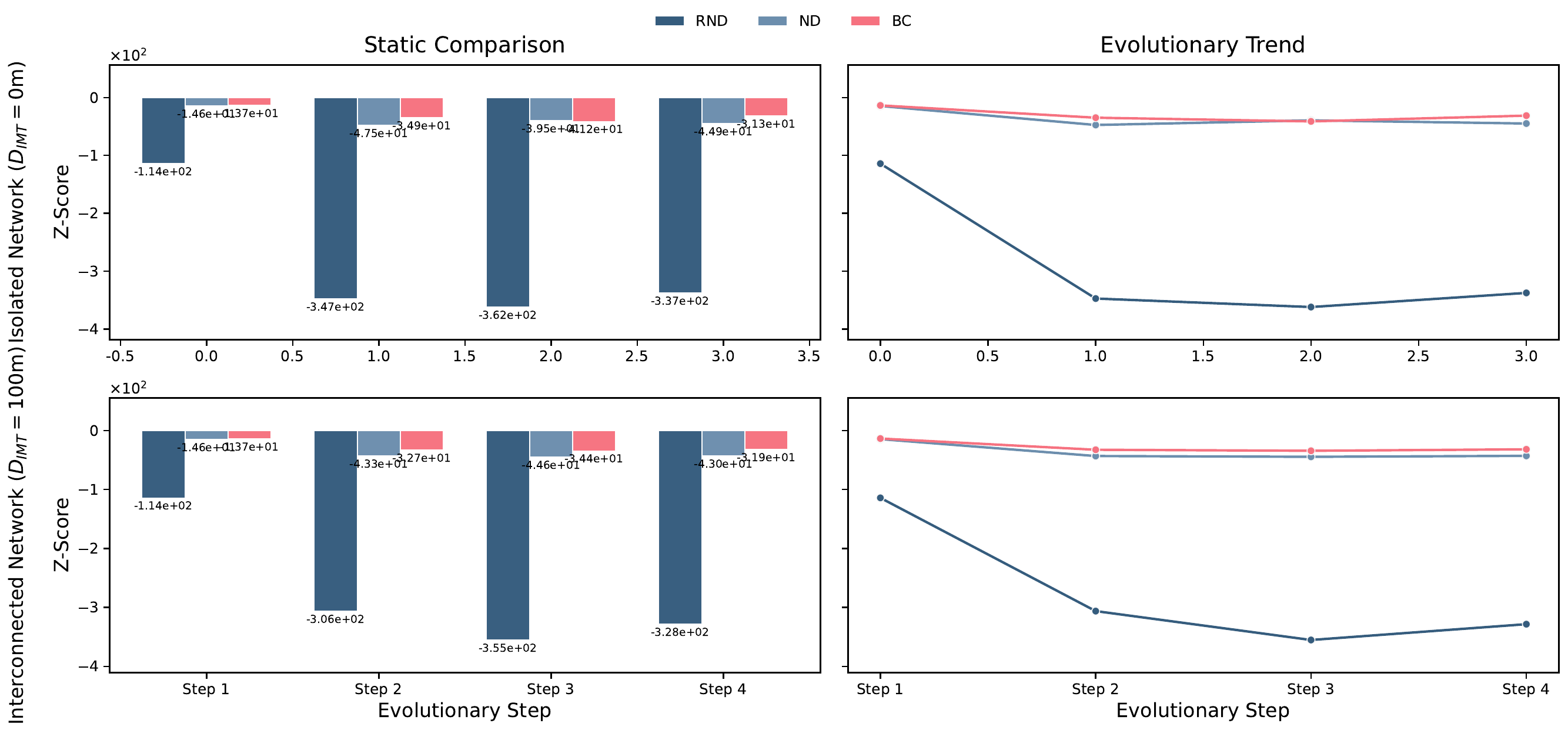}
    \caption{Z-score analysis of MPTN structural fragility. Panels compare isolated and interconnected configurations across integration steps and attack strategies.}
    \label{fig:z_core}
\end{figure*}

Benchmarking against geospatially constrained random graphs yields the same interpretation: Z-score patterns show that the integrated MPTN is structurally non-random across steps and attack strategies (Fig.~\ref{fig:z_core}). These diagnostics suggest that the static gains of integration reflect meaningful organization rather than artifacts of size or spatial embedding.

\begin{table}[htbp]
    \captionsetup{justification=raggedright, singlelinecheck=false, labelsep=colon, labelfont=bf, textfont=bf}
    \caption{Consolidated Relocation Rate ($R_\ell$) Analysis}
    \label{tab:consolidated_rl}
    \definecolor{tablegray}{HTML}{EAEBEB}
    \begin{tabularx}{\columnwidth}{l l *{4}{>{\raggedleft\arraybackslash}X}}
        \toprule
        \textbf{Subsystem} & \textbf{Model} & \multicolumn{2}{c}{$d_{\max}=750$\,m} & \multicolumn{2}{c}{$d_{\max}=1600$\,m} \\
        \cmidrule(lr){3-4} \cmidrule(lr){5-6}
        & & \multicolumn{1}{c}{\textbf{Iso.}} & \multicolumn{1}{c}{\textbf{Conn.}} & \multicolumn{1}{c}{\textbf{Iso.}} & \multicolumn{1}{c}{\textbf{Conn.}} \\
        \midrule
        \rowcolor{tablegray}
        & Symmetric & 0.0122 & 0.3934 & 0.3008 & 0.6019 \\
        \rowcolor{tablegray}
        \multirow{-2}{*}{Metro} & Asymmetric & 0.0122 & 0.3541 & 0.3008 & 0.5417 \\
        \addlinespace
        & Symmetric & 0.4881 & 0.4973 & 0.7483 & 0.7454 \\
        \multirow{-2}{*}{Bus} & Asymmetric & 0.4881 & 0.4881 & 0.7483 & 0.7483 \\
        \addlinespace
        \rowcolor{tablegray}
        & Symmetric & 0.0000 & 0.0000 & 0.0000 & 0.0000 \\
        \rowcolor{tablegray}
        \multirow{-2}{*}{Ferry} & Asymmetric & N/A & N/A & N/A & N/A \\
        \addlinespace
        & Symmetric & 0.0000 & 0.3235 & 0.0000 & 0.3235 \\
        \multirow{-2}{*}{Railway} & Asymmetric & N/A & N/A & N/A & N/A \\
        \bottomrule
    \end{tabularx}
    \begin{minipage}{\columnwidth}
    \vspace{\smallskipamount}
    \small
    Note: Iso. = Isolated, Conn. = Interconnected. "N/A" indicates the Asymmetric model is not applicable to single-mode systems.
    \end{minipage}
\end{table}

\subsection*{Dynamic cascade consequences}

\begin{table}[htbp]
\centering
\captionsetup{labelsep=colon, labelfont=bf, textfont=bf}
\caption{Dynamic cascade consequences across integration steps ($\beta=0.10$). Loss is the fraction of failed edges under node-triggered functional cascade simulations. FW/Sub denote the mean shares of failures from the first wave and subsequent overload cascade, respectively.}
\label{tab:expA_dynamic_consequences}
\scriptsize
\setlength{\tabcolsep}{3pt}
\renewcommand{\arraystretch}{1.0}
\resizebox{\columnwidth}{!}{%
\begin{tabular}{ll l r r r r r r r}
\toprule
Step & Trigger mode & Config & $n$ & Mean & P90 & Max & FW (\%) & Sub (\%) & Mean iters \\
\midrule
Step 1 & Metro & Iso. ($D_{IMT}=0$m) & 50 & 0.031 & 0.044 & 0.124 & 50.0 & 50.0 & 2.16 \\
       & Metro & Conn. ($D_{IMT}=100$m) & 50 & 0.032 & 0.067 & 0.160 & 52.2 & 47.8 & 2.20 \\
\midrule
Step 2 & Metro & Iso. ($D_{IMT}=0$m) & 50 & 0.131 & 0.133 & 0.135 & 0.5 & 99.5 & 10.00 \\
       & Metro & Conn. ($D_{IMT}=100$m) & 50 & 0.131 & 0.129 & 0.380 & 0.6 & 99.4 & 10.00 \\
       & Bus   & Iso. ($D_{IMT}=0$m) & 50 & 0.126 & 0.183 & 0.314 & 0.6 & 99.4 & 9.72 \\
       & Bus   & Conn. ($D_{IMT}=100$m) & 50 & 0.119 & 0.165 & 0.207 & 0.7 & 99.3 & 9.98 \\
\midrule
Step 3 & Metro & Iso. ($D_{IMT}=0$m) & 50 & 0.139 & 0.140 & 0.142 & 0.5 & 99.5 & 10.00 \\
       & Metro & Conn. ($D_{IMT}=100$m) & 50 & 0.128 & 0.149 & 0.365 & 0.6 & 99.4 & 9.98 \\
       & Bus   & Iso. ($D_{IMT}=0$m) & 50 & 0.130 & 0.143 & 0.182 & 0.6 & 99.4 & 9.96 \\
       & Bus   & Conn. ($D_{IMT}=100$m) & 50 & 0.126 & 0.160 & 0.345 & 0.7 & 99.3 & 9.84 \\
       & Ferry & Iso. ($D_{IMT}=0$m) & 3  & 0.116 & 0.117 & 0.117 & 0.4 & 99.6 & 10.00 \\
       & Ferry & Conn. ($D_{IMT}=100$m) & 3  & 0.093 & 0.093 & 0.093 & 0.4 & 99.6 & 10.00 \\
\midrule
Step 4 & Metro   & Iso. ($D_{IMT}=0$m) & 50 & 0.114 & 0.117 & 0.119 & 0.6 & 99.4 & 10.00 \\
       & Metro   & Conn. ($D_{IMT}=100$m) & 50 & 0.110 & 0.144 & 0.218 & 0.7 & 99.3 & 9.96 \\
       & Bus     & Iso. ($D_{IMT}=0$m) & 50 & 0.100 & 0.152 & 0.375 & 0.7 & 99.3 & 9.92 \\
       & Bus     & Conn. ($D_{IMT}=100$m) & 50 & 0.075 & 0.117 & 0.187 & 0.7 & 99.3 & 9.86 \\
       & Ferry   & Iso. ($D_{IMT}=0$m) & 3  & 0.028 & 0.028 & 0.028 & 1.1 & 98.9 & 6.00 \\
       & Ferry   & Conn. ($D_{IMT}=100$m) & 3  & 0.022 & 0.022 & 0.022 & 1.4 & 98.6 & 6.00 \\
       & Railway & Iso. ($D_{IMT}=0$m) & 3  & 0.028 & 0.028 & 0.028 & 1.1 & 98.9 & 6.00 \\
       & Railway & Conn. ($D_{IMT}=100$m) & 3  & 0.022 & 0.022 & 0.022 & 1.4 & 98.6 & 6.00 \\
\bottomrule
\end{tabular}%
}
\par\vspace*{3pt}
\noindent\begin{minipage}{\columnwidth}

\footnotesize
Note: Metro and Bus results are based on $n=50$ trigger nodes per configuration. Ferry and Railway results are based on $n=3$ trigger nodes in the study area and should be interpreted as indicative patterns rather than stable distributional estimates. Iso. = isolated; Conn. = interconnected.
\end{minipage}
\end{table}

We next examine functional consequences using node-triggered overload cascades. Step~1 shows short cascades, whereas from Step~2 onward losses are driven almost entirely by secondary overload. The effect of interconnection is mode-dependent: it generally reduces bus-triggered mean and extreme losses but amplifies metro-triggered extremes, especially in Steps~2 and~4. Static gains under integration therefore coexist with higher dynamic risk in specific failure modes (Table~\ref{tab:expA_dynamic_consequences}).

\begin{figure}[htbp]
    \centering
    \includegraphics[width=\columnwidth]{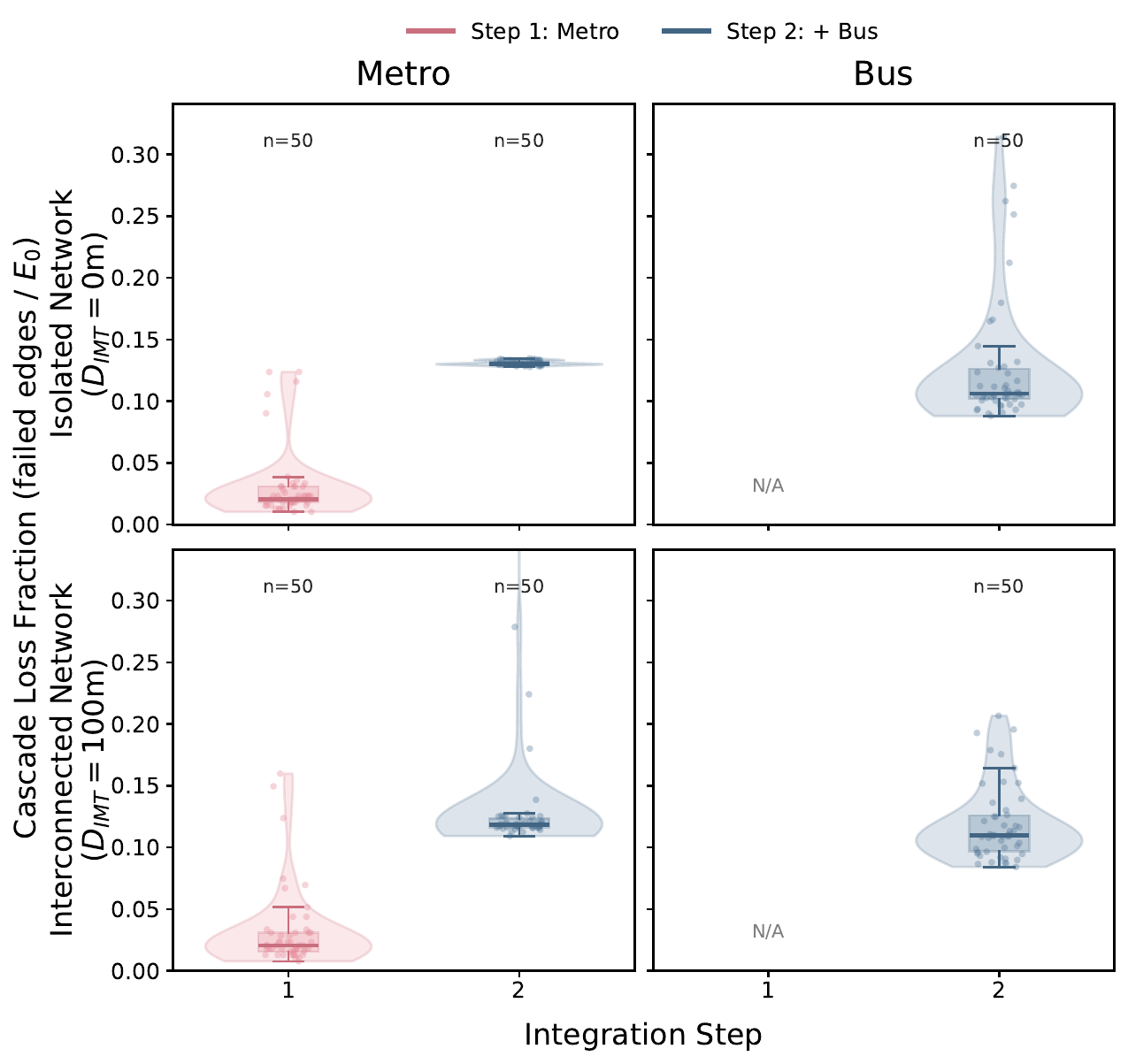}
    \caption{Distribution of cascade loss fraction ($|E_f|/E_0$) for Steps~1--2 under node-triggered functional cascade simulations. Rows compare isolated and interconnected configurations, and columns indicate metro- and bus-triggered failures. Losses rise sharply from Step~1 to Step~2; interconnection slightly reduces bus-triggered losses but amplifies the upper tail of metro-triggered losses.}
    \label{fig:step12}
\end{figure}

The distributional results confirm that the main transition occurs between Steps~1 and~2, when adding the bus layer sharply increases losses and turns a short response into an extended overload cascade (Fig.~\ref{fig:step12}). Mean cascade depth rises from about 2.2 iterations at Step~1 to nearly 10 from Step~2 onward, and more than 98\% of final damage comes from subsequent overload rather than the initial removal itself. Ferry- and railway-triggered results show similar but smaller losses, but they should be interpreted cautiously because only three trigger nodes are available in the study area.

\subsection*{Static descriptors have limited predictive power}

\begin{figure}[htbp]
    \centering
    \includegraphics[width=\columnwidth]{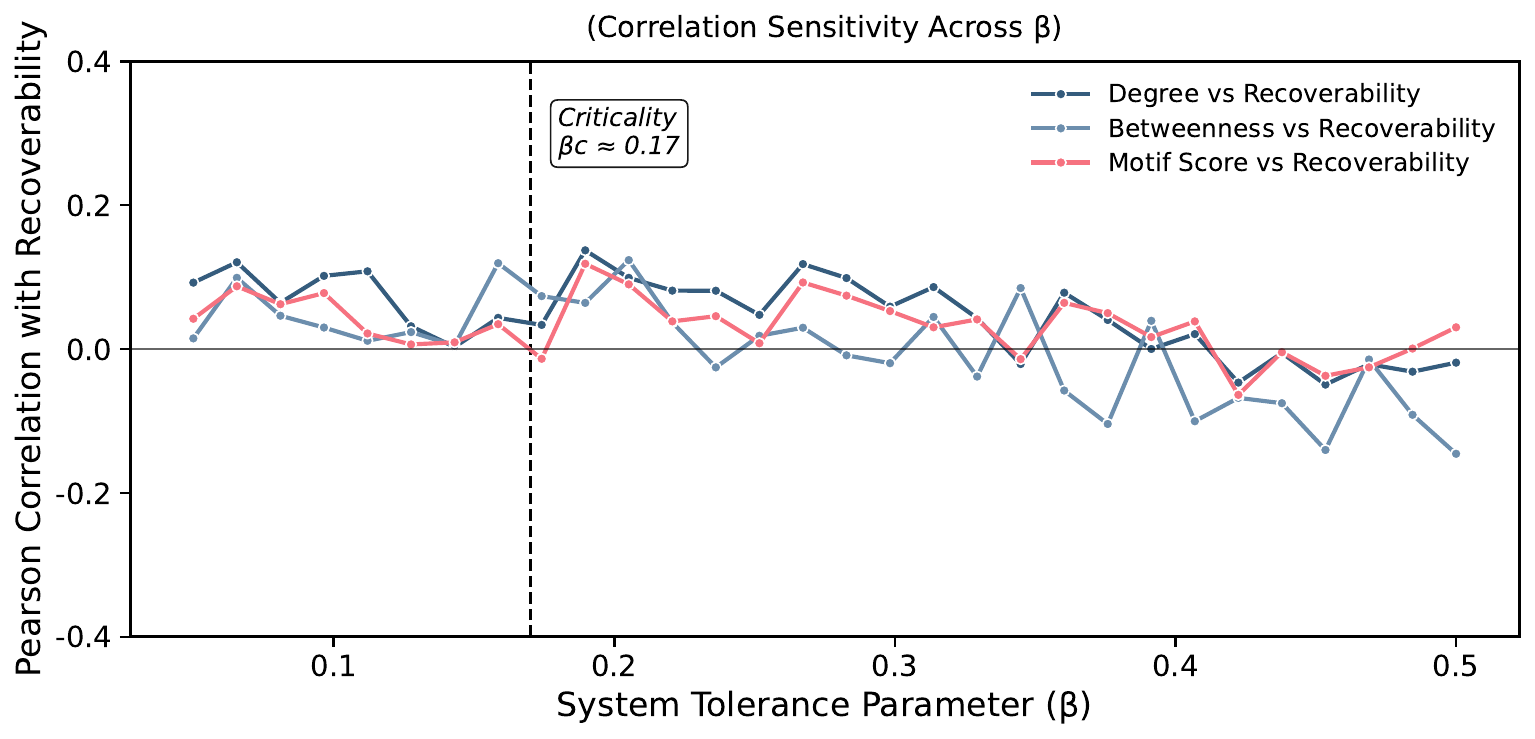}
    \caption{Correlations between static node metrics and functional recoverability across $\beta$. The dashed line marks the proxy-model critical region near $\beta_c \approx 0.17$; all associations remain weak over $\beta \in [0.05,0.5]$.}
    \label{fig:correlation}
\end{figure}

\begin{figure}[htbp]
    \centering
    \includegraphics[width=\columnwidth]{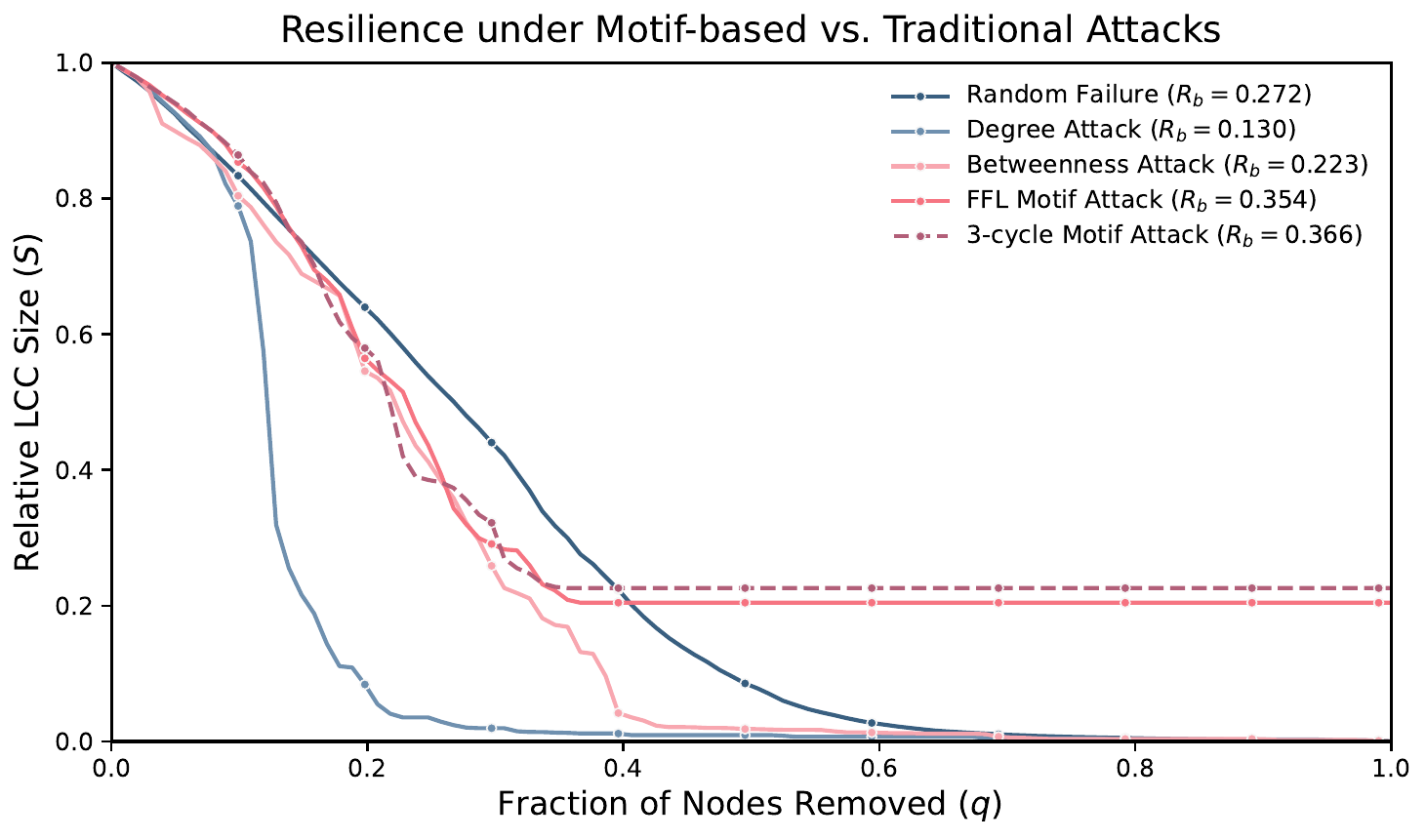}
    \caption{Network degradation under traditional and motif-based attacks. Motif-based rankings (FFL and directed 3-cycle) do not outperform degree- or betweenness-based attacks.}
    \label{fig:motif_attack}
\end{figure}

Static node descriptors provide limited guidance for functional recoverability. Across $\beta \in [0.05,0.5]$, correlations between recoverability and degree, betweenness, or motif participation remain weak and do not systematically increase above the proxy-model critical region near $\beta \approx 0.17$ (Fig.~\ref{fig:correlation}).

Rank-based tests lead to the same conclusion. Motif-based attacks derived from Feed-Forward Loops do not outperform degree- or betweenness-based attacks, and this result is unchanged when the motif is replaced by directed 3-cycles (Fig.~\ref{fig:motif_attack}). Static rankings therefore provide limited guidance for anticipating cascade damage.

\begin{figure*}[htbp]
    \centering
    \includegraphics[width=0.8\textwidth]{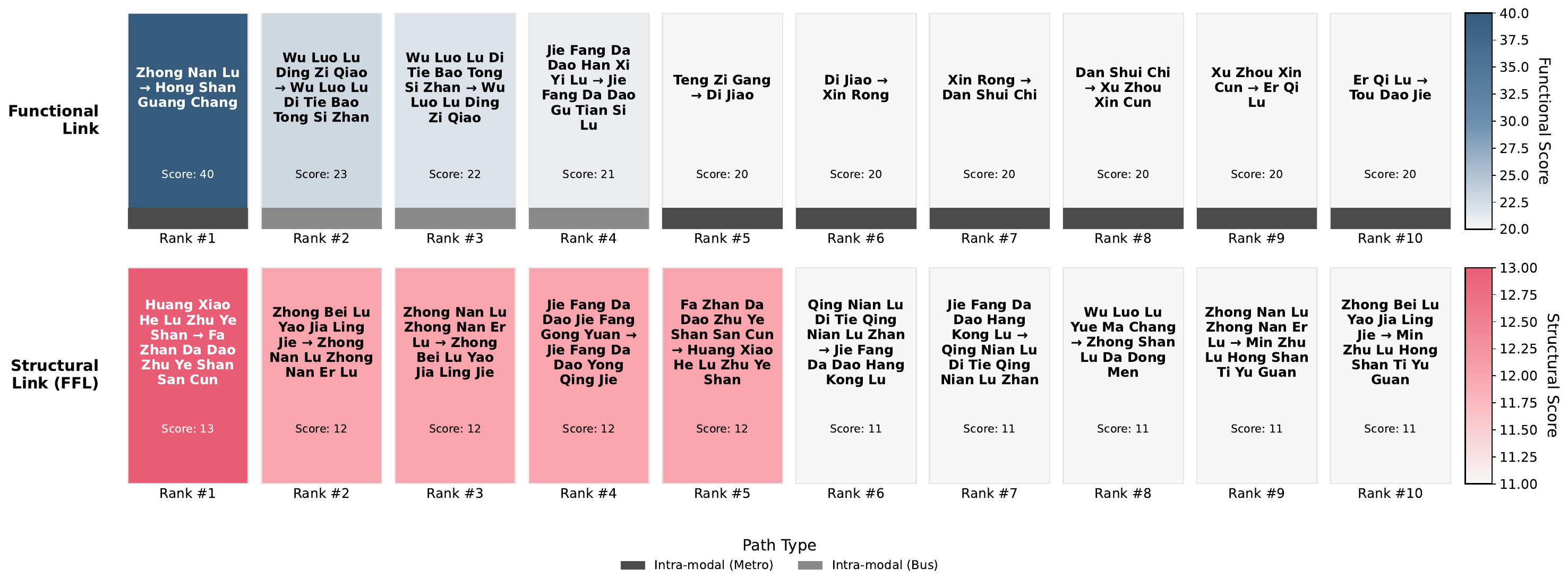}
    \caption{Functional and structural hierarchies in the MPTN. The upper panel ranks paths by weighted travel frequency, whereas the lower panel ranks them by Feed-Forward Loop (FFL) participation. Their mismatch highlights the weak alignment between functional use and motif-based structural prominence.}
    \label{fig:hierarchy_comparison}
\end{figure*}

This interpretation is consistent with the mismatch between functional and motif-based structural hierarchies. Frequently used paths are not those with the highest Feed-Forward Loop participation (Fig.~\ref{fig:hierarchy_comparison}), which helps explain the weak predictive value of static motif participation in this cascade setting.

\begin{figure}[htbp]
    \centering
    \includegraphics[width=\columnwidth]{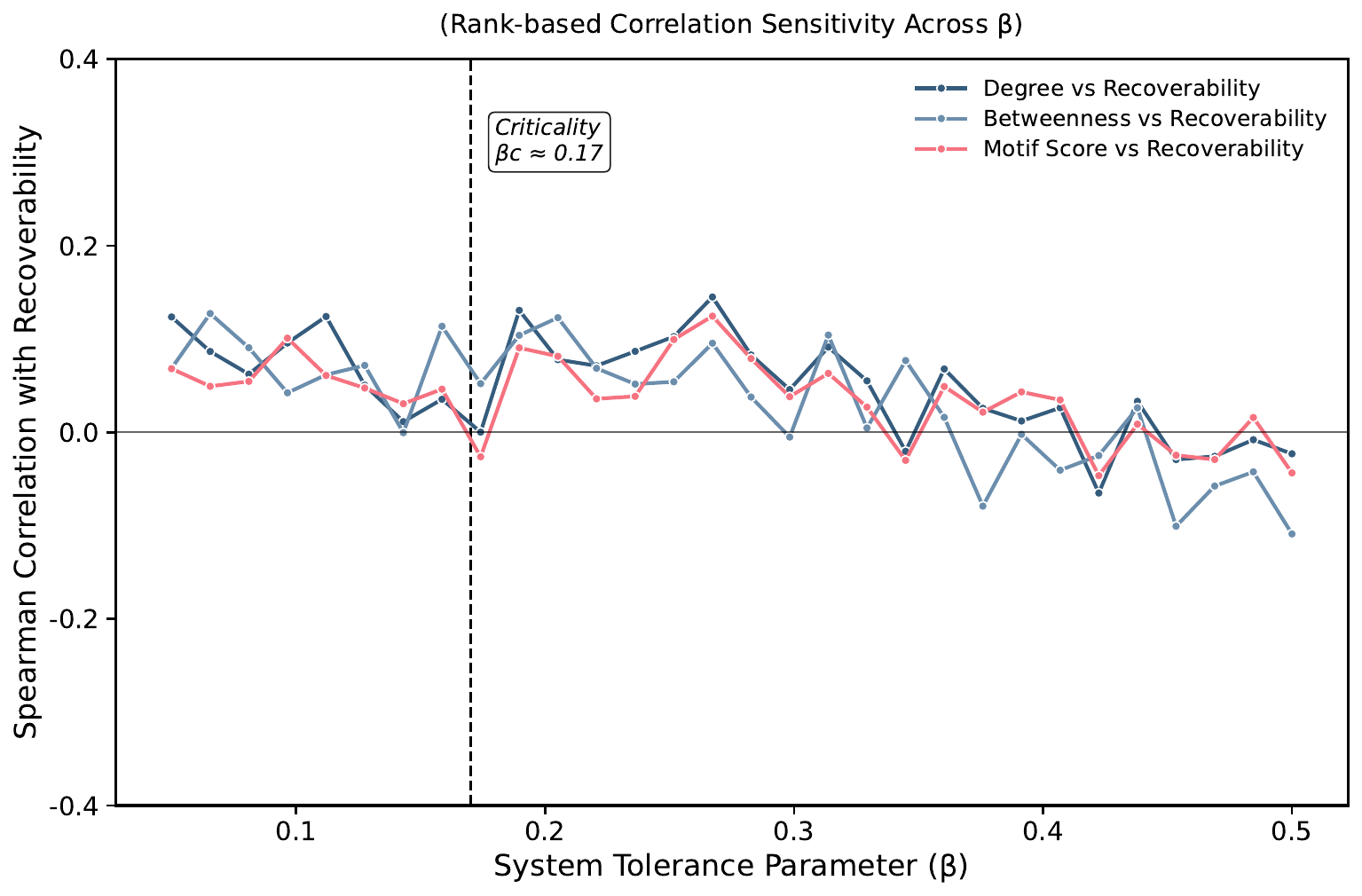}
    \caption{Spearman correlations between static node metrics and functional recoverability across $\beta \in [0.05,0.5]$. Correlations remain weak throughout.}
    \label{fig:spearman_correlation}
\end{figure}

A rank-based robustness check leads to the same conclusion. To complement the Pearson analysis in Fig.~\ref{fig:correlation}, we repeat the same test using Spearman rank correlation. The qualitative conclusion is unchanged: recoverability remains only weakly associated with degree, betweenness, and motif participation across the full tolerance range $\beta \in [0.05,0.5]$ (Fig.~\ref{fig:spearman_correlation}). This robustness check shows that the weak predictability of static node descriptors is not an artifact of using a linear correlation statistic.

\subsection*{Criticality and non-monotonic response}

\begin{figure}[htbp]
    \centering
    \resizebox{\columnwidth}{!}{%
        \includegraphics{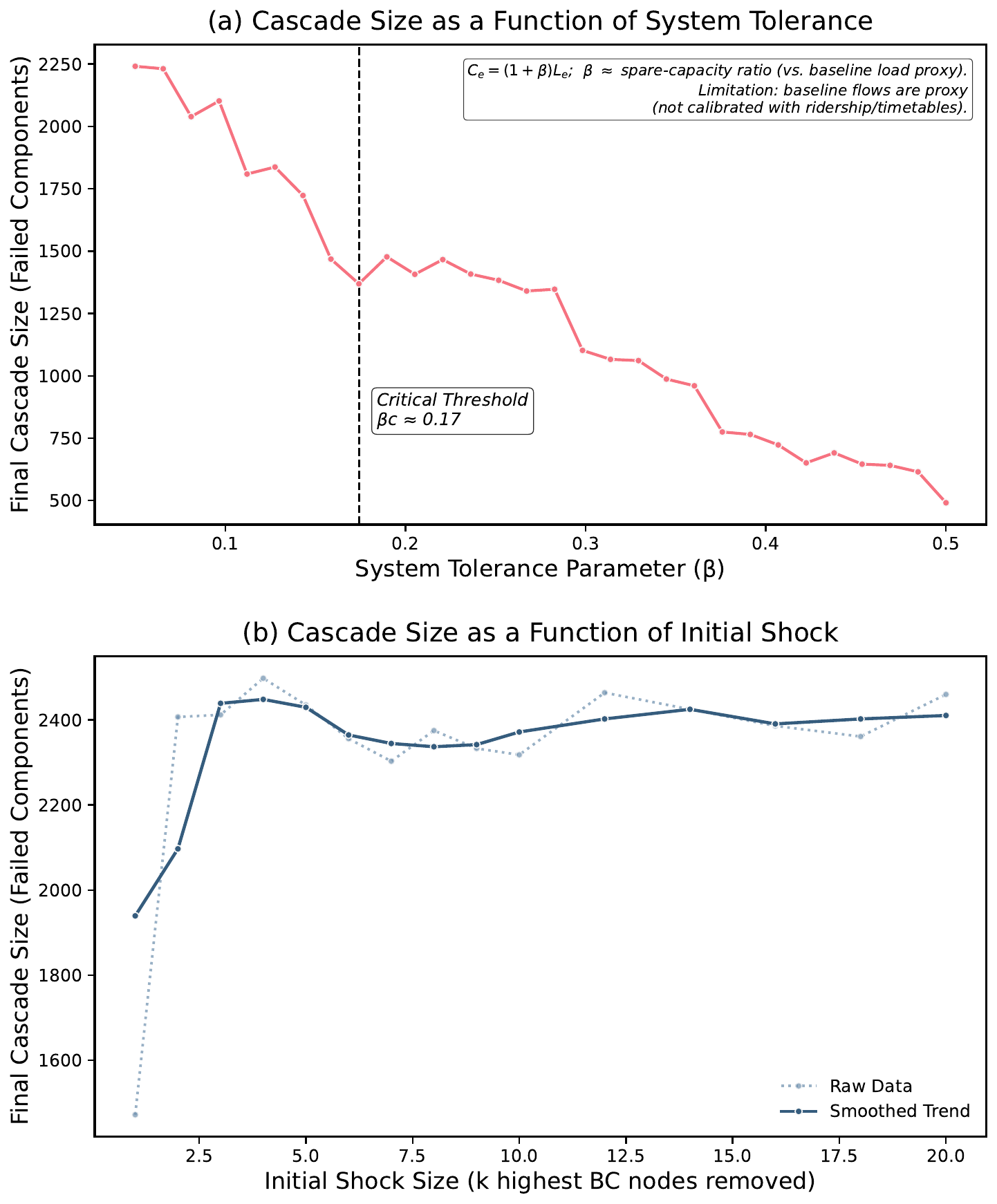}
    }
    
    \caption{Nonlinear dynamics of functional collapse. (a) Cascade size changes sharply as $\beta$ varies; the dashed line marks the proxy-model critical region near $\beta_c \approx 0.17$. Under $C_e=(1+\beta)L_e$, this value corresponds to about 17\% headroom relative to the proxy baseline load, or $L_e/C_e \approx 0.85$, and should therefore be interpreted as an internal stress marker rather than a calibrated operational spare-capacity threshold. (b) The response to increasing initial shock size ($k$) is non-monotonic.}
    \label{fig:nonlinear_dynamics}
\end{figure}

The cascade model exhibits clear system-level nonlinearity. Cascade size changes sharply around an apparent critical region near $\beta_c \approx 0.17$ in the normalized load--capacity setting. Under $C_e=(1+\beta)L_e$, this value corresponds to a proxy spare-capacity margin of about 17\% relative to baseline load, or equivalently a baseline utilization level of $(1+\beta_c)^{-1}\approx 0.85$. Because $L_e$ is estimated from shortest-path-based OD sampling rather than calibrated ridership or timetable flows, this threshold should not be interpreted as a measured reserve margin for Wuhan.

Nevertheless, its magnitude places the transition in a practically interpretable high-utilization regime. For comparison, a recent urban-rail capacity-control study examines train capacity-utilization limits of $1.0$, $0.9$, $0.8$, and $0.7$, while the \textit{Transit Capacity and Quality of Service Manual} treats vehicle loading, frequency, stations, terminals, and other operational constraints as central components of transit capacity assessment \cite{huang2024_toward,tcqsm2013}. Chinese urban-rail design guidance also requires no less than 10\% capacity reserve in design-period transport-capacity planning, which provides a domestic reference point while remaining distinct from the present proxy-load tolerance parameter \cite{gb55033_2022}. The transition near $\beta_c$ should therefore be read as an internal stress marker broadly comparable to high-utilization operating conditions, rather than as a calibrated spare-capacity requirement. This interpretation is consistent with prior cascade studies showing that tolerance-related parameters can govern abrupt transitions and that large passenger inflows can accelerate early-stage cascade spreading \cite{shen2022_capacity,gao2022_tl_rtn,lu2024_urtn} (Fig.~\ref{fig:nonlinear_dynamics}a).

The response to increasing initial shock size is also non-monotonic. Beyond a certain point, larger attacks can reduce final damage, consistent with a fragmentation or firewall-like effect (Fig.~\ref{fig:nonlinear_dynamics}b). These patterns explain why static structural importance does not translate reliably into dynamic failure consequences.

\section*{Discussion}

This study shows that intermodal integration in the Wuhan multimodal public transport network creates a tension between structural resilience and functional cascade risk. By combining safe-to-fail indicators, static structural descriptors, higher-order motif-based analysis and overload cascade simulation, we find that integration improves conventional static resilience indicators while producing mode-dependent effects under dynamic load redistribution. Metro--bus coupling increases connectivity, shortens paths, improves relocation opportunities and enhances robustness under standard attack-based diagnostics. These improvements indicate that intermodal integration strengthens the structural redundancy of the network.

The functional simulations, however, show that structural gains do not translate into uniformly safer cascade behavior. Interconnection generally reduces bus-triggered losses, suggesting that additional transfer opportunities can provide effective redistribution channels for some disruptions. At the same time, it can amplify the upper-tail risk of metro-triggered cascades, indicating that the same intermodal links that improve accessibility may also transmit overload to already important corridors. The main source of functional vulnerability in this setting is therefore not direct disconnection alone, but secondary overload generated by redistributed flows after disruption.

A central implication is that static structural importance is an incomplete proxy for functional recoverability in integrated multimodal transport networks. Degree, betweenness and motif participation all show weak associations with cascade outcomes across the tested tolerance range. The motif-based result should be interpreted carefully. It does not imply that higher-order approaches are irrelevant to transport resilience; rather, it shows that static triadic motif counts, when used only as node-ranking descriptors, do not capture the dynamic load-redistribution mechanism represented in the cascade model. Higher-order structure may still be important when group interactions, schedule coordination or passenger route-choice responses enter the propagation process directly.

These findings should be interpreted in light of the modelling choices. The load-capacity model uses shortest-path-based proxy loads rather than observed ridership, AFC records, timetable-based flows or station-level capacity measurements. Accordingly, the tolerance parameter $\beta$ represents a normalized stress condition in the simulation rather than a calibrated estimate of operational spare capacity. The apparent critical region near $\beta_c \approx 0.17$ should therefore be read as an internal stress marker of the proxy-load model, not as a validated reserve-margin threshold for the Wuhan system. This distinction is important because the value is useful for comparing simulated stress regimes, but it should not be used as a direct operational planning standard.

The analysis also treats the network as static. In practice, service frequency, schedule coordination, peak and off-peak demand, emergency control, and adaptive passenger behavior can reshape both rerouting opportunities and overload propagation. Inference for ferry- and railway-triggered cascades is also limited by the small number of trigger nodes available in the study area. These modelling boundaries do not undermine the main finding that static indicators provide limited guidance for dynamic cascade outcomes, but they define the conditions under which the finding should be generalized.

Future extensions should connect the present framework with observed passenger flows, AFC data, timetable information, station capacity estimates and behavior-aware route-choice models. Such extensions would allow the tolerance parameter to be related more closely to observable operating margins and would clarify how temporal service patterns and passenger adaptation influence cascade propagation. More generally, the results reposition multimodal transport resilience as a dynamic redistribution problem: integration can improve structural redundancy while still creating pathways through which local disruptions propagate into system-level functional losses.

\section*{Methods}

We study the Wuhan multimodal public transport network (MPTN) through an empirical pipeline that combines network construction, static structural characterization, higher-order motif analysis, overload cascade simulation and a stylized theoretical interpretation. The main text retains the assumptions, definitions and theoretical statements needed to interpret the empirical cascade simulations. Detailed derivations and proofs of the stylized theoretical results are provided in Supplementary Information.

\subsection*{Study area data and preprocessing}

Let $\mathcal{M}=\{\mathrm{metro},\mathrm{bus},\mathrm{ferry},\mathrm{rail}\}$ denote the set of transport modes. For each $m\in\mathcal{M}$, we use 2023 geospatial data, including district boundaries, station locations and route alignments, together with tabular route records containing ordered stop sequences.

The study area is the urban core of Wuhan, defined as the set of districts whose permanent-population urbanization rate exceeds $80\%$. Let $\mathcal{C}$ denote this district set and $\Omega_{\mathcal{C}}$ the merged urban-core polygon. Preprocessing removes invalid or temporary records, transforms all spatial data to WGS 84, harmonizes station records across sources and retains only stations and route segments intersecting $\Omega_{\mathcal{C}}$.

\begin{figure*}[htbp]
    \centering
    \includegraphics[width=0.85\textwidth]{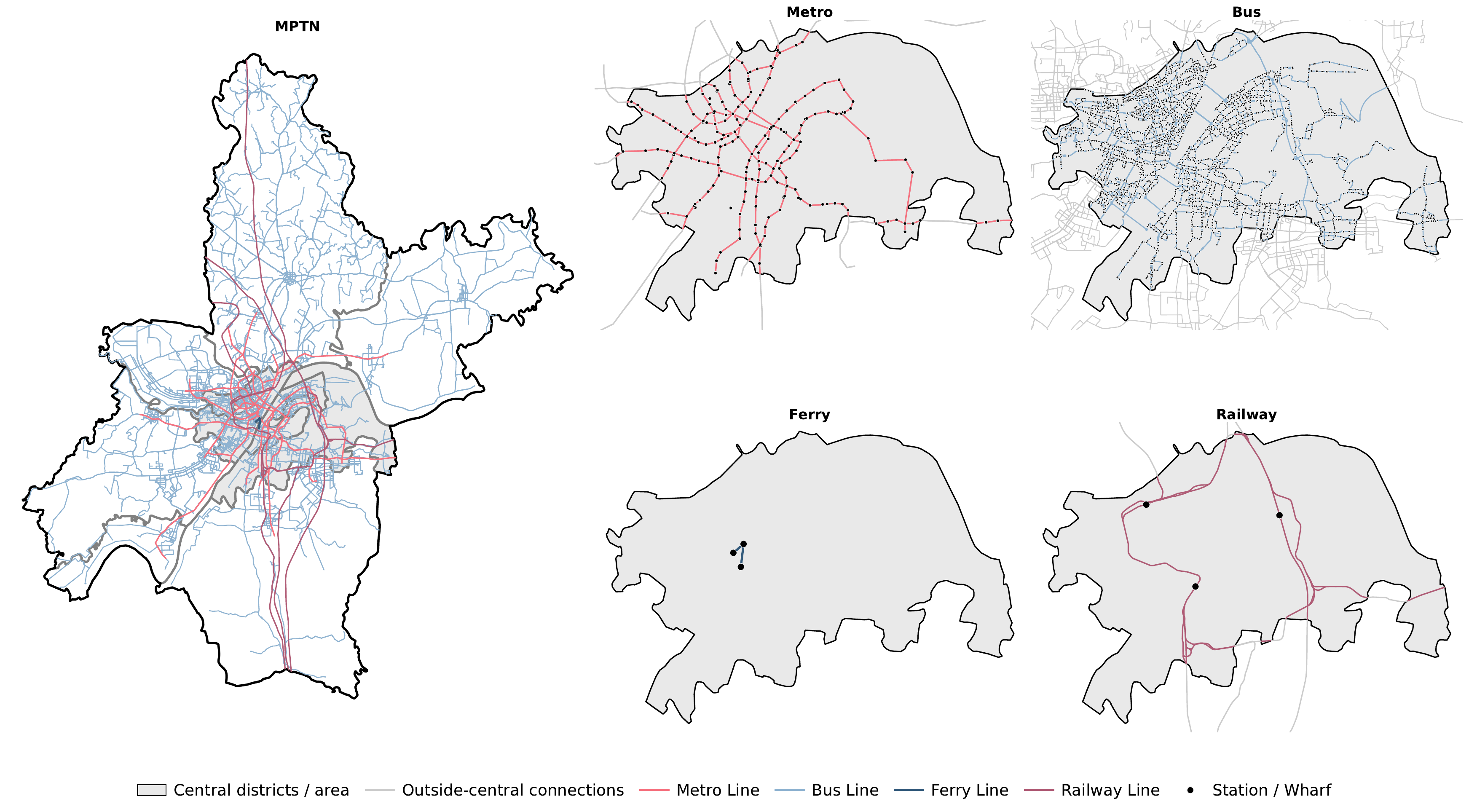}
    \caption{Wuhan MPTN in the study area, including the integrated network and four modal layers.}
    \label{fig:study_area_networks}
\end{figure*}

Figure~\ref{fig:study_area_networks} shows the processed study area network after spatial clipping and record harmonization, including the integrated MPTN and the four constituent modal layers retained for subsequent analysis.

\subsection*{Multilayer network construction}

We represent the MPTN as a directed multilayer graph
\begin{equation}
G=(V,E)=\Big(\bigcup_{m\in\mathcal{M}}V^{(m)},\; \bigcup_{m\in\mathcal{M}}E^{(m)} \cup E^{\mathrm{tr}}\Big),
\end{equation}
where $V^{(m)}$ and $E^{(m)}$ are the node and intra-modal edge sets of mode $m$, and $E^{\mathrm{tr}}$ is the set of inter-modal transfer edges.

Each node $v\in V^{(m)}$ represents one station of mode $m$ with a unique identifier, coordinates and mode label. For each operational route $r=(s_1,\dots,s_{n_r})$, we retain the maximal contiguous subsequence whose stations lie in $\Omega_{\mathcal{C}}$. For every retained subsequence, directed intra-modal edges $(s_k,s_{k+1})$ are added between consecutive stations and weighted by geographic distance.

To represent walkable transfers, we add a bidirectional inter-modal edge between $u\in V^{(m)}$ and $v\in V^{(m')}$, with $m\neq m'$, whenever
\begin{equation}
d(u,v)\le D_{\mathrm{IMT}},
\end{equation}
where $D_{\mathrm{IMT}}$ is the transfer threshold. Candidate transfer pairs are identified by k-d tree search. The resulting graph preserves directed service structure within each mode and explicit transfer accessibility across modes.

\subsection*{Static structural measures}

We use three static descriptors to characterize the pre-disruption resilience profile of the network: preparedness, robustness and interoperability.

Preparedness is quantified by the Gini coefficient of a node-level score $m:V\to\mathbb{R}_{+}$. Let $m_{(1)}\le \cdots \le m_{(N)}$, with $N=|V|$, denote the sorted values of $m$. We define
\begin{equation}
\mathrm{Gini}(m)=\frac{\sum_{i=1}^{N}(2i-N-1)m_{(i)}}{N^2\bar m},
\end{equation}
where $\bar m=\frac{1}{N}\sum_{i=1}^{N}m_{(i)}$. Smaller values indicate a more homogeneous distribution of node criticality.

For an attack rule $A$ and removal fraction $q\in[0,1]$, let $S_A(q)$ denote the relative size of the largest weakly connected component after removing a fraction $q$ of nodes. Structural robustness is defined as
\begin{equation}
r_b(A)=\int_0^1 S_A(q)\,dq.
\end{equation}

Interoperability is measured by the relocation rate. For a disrupted node $v$, let $\mathrm{desc}(v)$ denote the set of nodes reachable from $v$ in the original network. For each $n\in\mathrm{desc}(v)$, let $u_n^*$ be the neighbor of $v$ that yields the shortest alternative path to $n$ after the disruption of $v$. The node-level relocation rate is
\begin{equation}
R_\ell(v)=\frac{1}{|\mathrm{desc}(v)|}\sum_{n\in\mathrm{desc}(v)}
\left(1-\frac{d(v,u_n^*)}{d_{\max}}\right),
\end{equation}
where $d_{\max}$ is the relocation distance limit. The network-level relocation rate is
\begin{equation}
R_\ell=\frac{1}{|V'|}\sum_{v\in V'}R_\ell(v),
\end{equation}
where $V'\subseteq V$ is the set of evaluated disrupted nodes.

\subsection*{Higher-order descriptors and motif-based attacks}

To test whether static higher-order structure improves vulnerability assessment, we assign each node a motif participation score based on directed triadic motifs.

Our primary motif is the Feed-Forward Loop (FFL), interpreted here as a triadic bypass unit: a direct corridor $u\to w$ together with a two-step bypass $u\to v\to w$. We focus on the FFL because it is the smallest directed motif that matches the local rerouting mechanism in the overload model. When the direct corridor is disrupted, load may be redistributed to the two-step bypass and induce secondary overload. This mechanism is summarized below through the triadic bypass unit and the single-shock capacity-feasibility criterion. Thus, the FFL is used as a minimal static proxy for rerouting-driven local dependency. For each node, the FFL score is the number of such motifs in which it participates.

As a robustness check, we also consider directed 3-cycles, which capture closed directed triadic connectivity but do not represent the same direct-corridor-versus-bypass mechanism. For each motif type, nodes are ranked by participation count, and the resulting ranking defines a motif-based targeted attack.

\subsection*{Dynamic cascade simulation and recoverability}

Functional resilience is evaluated using an overload-driven cascade model. Baseline edge loads $L_e$ are estimated by sampled origin--destination assignments along shortest paths. Edge capacities are defined as
\begin{equation}
C_e=(1+\beta)L_e,
\end{equation}
where $\beta>0$ is a dimensionless tolerance parameter.

Because $L_e$ is derived from shortest-path OD sampling rather than calibrated ridership or timetable data, $\beta$ is interpreted as a normalized stress parameter rather than a direct operational spare-capacity percentage. This is consistent with prior metro cascading-failure studies in which tolerance-type parameters are analytically useful but not directly calibrated to observed operating reserve \cite{shen2022_capacity}. Larger $\beta$ indicates more residual headroom against rerouted overload, whereas smaller $\beta$ corresponds to a tighter operating regime. We therefore vary $\beta$ to test whether the weak association between static structure and functional recoverability is stable across stress conditions, rather than to infer a calibrated operating threshold for the Wuhan system.

Given an initial node disruption, flows are reassigned on the residual network. Any edge whose updated load exceeds its capacity fails, and the process iterates until no new overload failures occur. For a cascade triggered at node $v$, functional recoverability is defined as
\begin{equation}
R_{\mathrm{recover}}(v)=1-\frac{|E_{\mathrm{failed}}|}{|E_{\mathrm{initial}}|}.
\end{equation}

\subsection*{Statistical protocols}

Random-failure simulations are repeated 50 times to stabilize stochastic estimates. For null-model benchmarking, we generate 50 Erd\H{o}s--R\'enyi random graphs preserving the number of nodes, number of edges and node geolocations. For a statistic $X$, the corresponding null-model score is
\begin{equation}
Z=\frac{X_{\mathrm{real}}-\mu_{\mathrm{rand}}}{\sigma_{\mathrm{rand}}},
\end{equation}
where $\mu_{\mathrm{rand}}$ and $\sigma_{\mathrm{rand}}$ are the mean and standard deviation of the reference ensemble.

\subsection*{Stylized theoretical interpretation}
\label{sec:stylized_theory}

This subsection provides a stylized interpretation of the overload cascade simulator rather than a calibrated mathematical model of the Wuhan system. It retains the main theoretical statements needed to interpret the empirical results: the single-shock overload criterion, the triadic bypass implication of the Feed-Forward Loop motif, and the branching-process explanation of tolerance-dependent amplification. Formal auxiliary results, including cascade-operator monotonicity, finite termination, the static-betweenness counterexample, and all proofs are provided in Supplementary Information.

Let $G=(V,E)$ be a finite directed graph, where nodes denote transport facilities and directed edges denote service links. An origin--destination demand matrix induces a baseline edge-load vector $L^{(0)}$. For a tolerance parameter $\beta>0$, edge capacity is defined as
\begin{equation}
C_e=(1+\beta)L_e^{(0)}.
\end{equation}
For a failure set $F\subseteq E$, let $L^{(F)}$ denote the rerouted load vector on the residual network. A surviving edge fails when its rerouted load exceeds capacity, which defines the cascade operator
\begin{equation}
\Phi(F)
=
F\cup
\left\{
e\in E\setminus F:
L_e^{(F)}>C_e
\right\}.
\end{equation}
The cascade trajectory is $F_{t+1}=\Phi(F_t)$. Because the graph is finite, and because overload failures are accumulated monotonically in the simulator, the process reaches a fixed point after finitely many iterations. This provides the stopping rule for the empirical cascade simulation.

For a single-edge initial failure $F=\{e^\star\}$, suppose that a fraction $\Gamma_{e^\star\to e}$ of the disrupted load on $e^\star$ is redistributed to a surviving edge $e$.

\begin{lemma}[Single-shock capacity-feasibility criterion]
\label{lem:single_shock_main}
The receiving edge $e$ avoids overload if and only if
\begin{equation}
\Gamma_{e^\star\to e}L_{e^\star}^{(0)}
\le
\beta L_e^{(0)}.
\end{equation}
Equivalently, defining
\begin{equation}
\Theta_{e^\star\to e}
=
\frac{
\Gamma_{e^\star\to e}L_{e^\star}^{(0)}
}{
L_e^{(0)}
},
\end{equation}
the no-overload condition is
\begin{equation}
\Theta_{e^\star\to e}\le\beta.
\end{equation}
\end{lemma}

This criterion is the local mechanism behind redistribution-driven secondary failure. It shows that overload depends on disrupted load, redistribution share, receiving-edge baseline load, and tolerance. These quantities are not determined by static topology alone, which explains why degree, betweenness, and motif participation may provide only weak guidance for functional recoverability.

The same criterion gives a direct interpretation of the Feed-Forward Loop motif. We treat the FFL as a triadic bypass unit consisting of a direct corridor $e_3=(u,w)$ and a two-segment bypass $e_1=(u,v)$, $e_2=(v,w)$.

\begin{corollary}[Triadic bypass overload condition]
\label{cor:tbu_main}
Suppose that the direct edge $e_3=(u,w)$ fails and that its disrupted load is rerouted entirely through the two-segment bypass, so that $\Gamma_{e_3\to e_2}=1$. Then the downstream bypass edge $e_2=(v,w)$ fails at the next cascade step if and only if
\begin{equation}
L_{e_3}^{(0)}
>
\beta L_{e_2}^{(0)}.
\end{equation}
\end{corollary}

Corollary~\ref{cor:tbu_main} explains why FFL participation is a meaningful but incomplete static descriptor. The motif identifies a possible local rerouting structure, but whether that structure absorbs or transmits overload depends on load ratios and available tolerance rather than on motif count alone.

At the system level, let $\Theta$ denote a generic overload ratio and let $k_{\mathrm{eff}}$ denote the effective number of downstream recipient edges exposed by one failed edge. Define
\begin{equation}
\lambda(\beta)
=
\mathbb{E}[k_{\mathrm{eff}}]\,
\mathbb{P}(\Theta>\beta).
\end{equation}
Under a Galton--Watson approximation, $\lambda(\beta)$ is the expected number of secondary failures generated by one failed edge.

\begin{theorem}[Expected cascade size under a Galton--Watson approximation]
\label{thm:expected_size_main}
Assume that the stylized cascade is approximated by a Galton--Watson branching process with offspring mean $\lambda(\beta)$, and let $Z$ denote the total progeny including the initial failed edge.

\textbf{(i) Subcritical regime.} If $\lambda(\beta)<1$, then
\begin{equation}
\mathbb{E}[Z]
=
\frac{1}{1-\lambda(\beta)},
\qquad
\mathbb{E}[Z-1]
=
\frac{\lambda(\beta)}{1-\lambda(\beta)}.
\end{equation}

\textbf{(ii) Critical and supercritical regimes.} If $\lambda(\beta)\ge1$, then
\begin{equation}
\mathbb{E}[Z]=+\infty
\end{equation}
in the idealized unbounded branching process. If $\lambda(\beta)>1$, the extinction probability $q$ is the smallest solution of $f(q)=q$, and the survival probability is $1-q>0$.

\textbf{(iii) Finite-system truncation.} For $Z_M=\min\{Z,M\}$ and $\lambda(\beta)>1$,
\begin{equation}
(1-q)M
\le
\mathbb{E}[Z_M]
\le
M.
\end{equation}
\end{theorem}

Theorem~\ref{thm:expected_size_main} provides the aggregate interpretation of the nonlinear transition observed in the simulations. When tolerance is high or redistribution exposure is limited, secondary failures decay on average. When tolerance decreases or more downstream edges are exposed, $\lambda(\beta)$ can approach or exceed unity, producing abrupt amplification in the idealized process. In the finite empirical network, cascades still terminate, but the branching approximation explains why cascade size can change sharply as $\beta$ varies.

Together, these stylized results support the main empirical interpretation of the study. Cascades terminate because the network is finite; local bypasses can either absorb or transmit overload depending on capacity feasibility; FFL motifs identify rerouting opportunities but not their functional safety; and tolerance-dependent amplification explains the observed nonlinear cascade response. Static structural descriptors, including higher-order motif scores, are therefore insufficient proxies for functional recoverability when cascade outcomes are governed by dynamic load redistribution.

\section*{Data availability.} 
Data were collected from the National Platform for Common GeoSpatial Information Services \url{https://www.tianditu.gov.cn/} and Baidu Map using a web crawler \url{https://map.baidu.com/}. The dataset pertains to the year 2023. The railway subsystem includes passenger high-speed rail and intercity rail; stations that do not handle operational passenger train services were excluded from the analysis.

\section*{Code availability.}
The codes used for data processing and analysis are available via GitHub: \url{https://github.com/yzz980314/high-order-network-in-city.git}.

\bibliographystyle{IEEEtran}
\bibliography{references}

\section*{Acknowledgements.}
We gratefully acknowledge the National Natural Science Foundation of China (No. 50608061) and the Graduate Innovation Project of National University of Defense Technology (Grant No. XJQY2024065) for funding this research. The funders had no role in the design of the study, the collection, analysis, or interpretation of data, the writing of the manuscript, or the decision to submit the article for publication.
\section*{Author Contributions.}
J.S. conceived and designed the experiments. Y.W. performed the experiments and analyzed the data. Z.Y. contributed materials/analysis tools. Z.Y. and Y.W. wrote the paper. All authors reviewed the manuscript.

\section*{Competing Interests.}
The authors declare no competing interests.

\section*{Additional information}
\textbf{Correspondence} and requests for materials should be addressed to Zimo Yan.

\clearpage
\noindent\textbf{Figure titles and legends}

\par\medskip
\noindent\textbf{Fig. 1: Analytical framework for the Wuhan MPTN}
Analytical framework for the Wuhan MPTN. (a) Multilayer network construction. (b) Traditional structural analysis, higher order motif analysis, and dynamic cascade simulation. (c) Synthesis of the main findings: integration improves static resilience indicators, while functional cascade outcomes depend on dynamic redistribution and are weakly predicted by static node descriptors.

\par\medskip

\noindent\textbf{Fig. 2: Robustness of the MPTN across integration steps}

Robustness of the MPTN across integration steps under isolated ($D_{\mathrm{IMT}}=0$ m) and interconnected ($D_{\mathrm{IMT}}=100$ m) configurations.

\par\medskip
\noindent\textbf{Fig. 3: Z-score analysis of MPTN structural fragility}

Z-score analysis of MPTN structural fragility. Panels compare isolated and interconnected configurations across integration steps and attack strategies.

\par\medskip
\noindent\textbf{Fig. 4: Distribution of cascade loss fraction for Steps 1--2}

Distribution of cascade loss fraction ($|E_f|/E_0$) for Steps~1--2 under node-triggered functional cascade simulations. Rows compare isolated and interconnected configurations, and columns indicate metro- and bus-triggered failures. Losses rise sharply from Step~1 to Step~2; interconnection slightly reduces bus-triggered losses but amplifies the upper tail of metro-triggered losses.

\par\medskip
\noindent\textbf{Fig. 5: Correlations between static node metrics and functional recoverability}

Correlations between static node metrics and functional recoverability across $\beta$. The dashed line marks the proxy-model critical region near $\beta_c \approx 0.17$; all associations remain weak over $\beta \in [0.05,0.5]$.

\par\medskip
\noindent\textbf{Fig. 6: Network degradation under traditional and motif-based attacks}

Network degradation under traditional and motif-based attacks. Motif-based rankings (FFL and directed 3-cycle) do not outperform degree- or betweenness-based attacks.

\par\medskip
\noindent\textbf{Fig. 7: Functional and structural hierarchies in the MPTN}

Functional and structural hierarchies in the MPTN. The upper panel ranks paths by weighted travel frequency, whereas the lower panel ranks them by Feed-Forward Loop (FFL) participation. Their mismatch highlights the weak alignment between functional use and motif-based structural prominence.

\par\medskip
\noindent\textbf{Fig. 8: Spearman correlations between static node metrics and functional recoverability}

Spearman correlations between static node metrics and functional recoverability across $\beta \in [0.05,0.5]$. Correlations remain weak throughout.

\par\medskip
\noindent\textbf{Fig. 9: Nonlinear dynamics of functional collapse}
Nonlinear dynamics of functional collapse. (a) Cascade size changes sharply as $\beta$ varies; the dashed line marks the proxy-model critical region near $\beta_c \approx 0.17$. Under $C_e=(1+\beta)L_e$, this value corresponds to about 17\% headroom relative to the proxy baseline load, or $L_e/C_e \approx 0.85$, and should therefore be interpreted as an internal stress marker rather than a calibrated operational spare-capacity threshold. (b) The response to increasing initial shock size ($k$) is non-monotonic.
\par\medskip
\noindent\textbf{Fig. 10: Wuhan MPTN in the study area}

Wuhan MPTN in the study area, including the integrated network and four modal layers.

\end{document}